\documentclass[12pt,a4paper]{article}

\usepackage{geometry}
\usepackage{amsmath}
\usepackage{amssymb}
\usepackage[dvips]{graphicx}
\usepackage{cite}
\usepackage{pstricks}
\usepackage{bm}
\usepackage{pbox}
\usepackage{placeins}
\usepackage{graphicx}
\usepackage{caption}
\usepackage{subcaption}
\usepackage[T1]{fontenc}
\usepackage{footnote}
\usepackage{pdfpages}
\usepackage{hhline}
\usepackage{multirow}
\usepackage{feynmf}
\usepackage{multicol}
\usepackage{enumitem}
\usepackage{multirow}
\usepackage{relsize}
\usepackage{slashed}
\usepackage{graphicx}
\usepackage{lineno,hyperref}
\usepackage{comment}
\modulolinenumbers[5]
\usepackage[toc,page]{appendix}
\allowdisplaybreaks

\usepackage{slashed}
\usepackage{cases}

\usepackage{empheq}

\definecolor{MyDarkBlue}{rgb}{0.1, 0.1, 0.8} 
\definecolor{SBlue}{rgb}{0.2, 0.4, 0.7} 
\definecolor{MyLightBlue}{rgb}{0.22,0.51,0.9}
\definecolor{MyGreen}{rgb}{0.0, 0.5, 0.0}
\definecolor{BrickRed}{rgb}{0.8, 0.25, 0.33}
\RequirePackage{hyperref}
\hypersetup{colorlinks, citecolor=SBlue,linkcolor=MyDarkBlue, urlcolor=BrickRed}

\begin{document}
\vspace*{-0.2in}
\begin{flushright}
\end{flushright}
\vspace{0.1cm}
\begin{center}
{\Large \bf
Bounding exotic top decays
inclusively \\ 
\vskip 0.3cm 
at the FCC-ee
}
\end{center}
\renewcommand{\thefootnote}{\fnsymbol{footnote}}
\vspace{0.8cm}
\begin{center}
{
{}~\textbf{Gennaro Corcella$^1$}\footnote{ E-mail: \textcolor{MyDarkBlue}{Gennaro.Corcella@lnf.infn.it}},
{}~\textbf{Barbara Mele$^2$}\footnote{ E-mail: \textcolor{MyDarkBlue}{barbara.mele@roma1.infn.it}},
{}~\textbf{Dibyashree Sengupta$^{2}$}\footnote{ E-mail: \textcolor{MyDarkBlue}{dibyashree.sengupta@roma1.infn.it}}
}
\vspace{0.5cm}
{
\\
\em $^1$ INFN, Laboratori Nazionali di Frascati,
Via E.~Fermi 54, I-00044 Frascati (RM), Italy
\\
$^2$ INFN, Sezione di Roma,  
c/o Dip. di Fisica, Sapienza Università di Roma, \\
Piazzale Aldo Moro 2, I-00185 Rome, Italy
} 
\end{center}

\renewcommand{\thefootnote}{\arabic{footnote}}
\setcounter{footnote}{0}
\thispagestyle{empty}

\vspace{0.8cm}
\begin{abstract}
Since its discovery, the top quark has never been
produced and studied 
in an environment as clean as that predicted for
$e^+e^-$ collisions at future colliders.
Details of the top quark's properties, completely unattainable 
in hadronic collisions, can be analyzed via lepton collisions.
New strategies for analyzing the physics of the top 
quark can, therefore, be developed in such a 
spectacularly clean environment.
Here we focus on the possibility of inclusively measuring exotic excesses 
in the top decay width by studying
the direct production of 
$t\bar t$  at the FCC-$ee$, thus establishing model-independent 
limits for rare decays branching fractions of the top quark.
\end{abstract}

\setcounter{footnote}{0}

\newpage


\section{Introduction}
\label{sec:intro}

High-energy lepton collisions in machines such as the FCC-$ee$~\cite{FCC:2025lpp, FCC:2025uan, FCC:2025jtd}, CEPC~\cite{CEPCStudyGroup:2023quu}, ILC~\cite{Bambade:2019fyw}
or  CLIC~\cite{CLICdp:2018cto} offer possibilities for exploring precision physics that may be well beyond the reach of
hadron accelerators such as the Large Hadron Collider (LHC) at CERN, even though the latter is characterized by larger centre-of-mass (c.o.m.) energies.
Although hadron colliders can, in general, yield much larger production cross sections and available phase space, and thus have a
greater potential for the direct discovery  of new heavy states, electron-positron collisions benefit from a remarkable democracy in production cross sections, all of which are of electroweak origin. 
Consequently, although production statistics are generally moderate in $e^+e^-$ colliders, 
it is possible to exploit, on the one hand, very accurate theoretical predictions and, on the other,
a clean experimental environment with smaller backgrounds than in hadron collisions.

As a result,  $e^+e^-$ colliders 
 can operate in untriggered mode and, in principle,
detect and reconstruct any ``unexpected'' final state, including hadronic final states
or what might be {\it invisible} at the LHC simply because it is not known  {\it a priori} what
to trigger on.

One of the physics sector which would extensively benefit from this situation is top-quark phenomenology~\footnote{A review on top quark studies at the HL-LHC and future colliders can be found in~\cite{Schwienhorst:2022yqu}.}.
While top-pair production in $e^+e^-$ collisions at threshold has been thoroughly explored  (for a review see~\cite{Defranchis:2025auz}),
a somewhat less extensive investigation has been undertaken for
top production in the continuum, well above the top pair threshold. 
As $e^+e^-$ collisions will allow one to trace back top-quark final states
almost on an event-by-event basis, one will actually be able to look at
details of top production and kinematics that are not even conceivable in hadron collisions. Relevant strategies are mostly yet to be developed.

Rare decays are one of the  top physics chapters that are expected
to widely benefit from the spectacularly clean environment of $e^+e^-$ collisions. In this paper we wish
to focus on the possibility to bound {\it exotic} contributions 
to the top decay width in an inclusive (i.e. model independent) manner,
which can be alternative to the top width determination through
the cross section measurement at threshold~\cite{Defranchis:2025auz}.

The heavy mass of the top quark might, in principle, allow it to decay into many different ``unpredicted'' final states with correspondingly unknown kinematical features (see e.g.~\cite{Antusch:2025lpm}). For instance, in different Beyond-Standard-Model (BSM) frameworks, one could 
have the decay channels : $i)\; t \to b H^+ \to \tau\nu b$,
$H^+$ being a charged Higgs boson;
$ii) \;t \to Z'c, Z'u,$, where $Z'$ is a light non-standard gauge boson; 
$iii) \;t \to \chi\chi c, \chi\chi u\;$, $\chi$ being a Dark Matter candidate; 
$iv) \;t \to n$-jets, where the jets are not mediated by a $b$ quark and a $W$ boson. 
In all
of the above cases, the  
 final-state composition and kinematics do not match the Standard Model (SM) dominant top decay channel $t\to Wb$. 

Note that, unless one assumes a particular BSM top decay pattern, LHC experiments are not able to select experimentally a generic exotic top decay due to critical backgrounds. On the other hand, thanks also to the closed kinematics of the $e^+e^-\to t\,\bar{t}\,$ products, once a SM top quark is tagged in the final state, one can, in principle, study the recoil system in an inclusive way, independently of the actual second-top decay channel.
This strategy is inspired by  Higgs studies at $e^+e^-$ colliders. Indeed, in the associated production $e^+e^-\to ZH$, one can determine some Higgs properties by just measuring the $Z$ decay products and reconstructing the $Z$ recoil system in an inclusive ({\it blind}) way, irrespective
of the particular Higgs decay mode~\cite{Ito:2009jf, Lohmann:2007ty}. 

\begin{figure}[htbp]
\begin{center}
  \includegraphics[height=0.15\textheight]{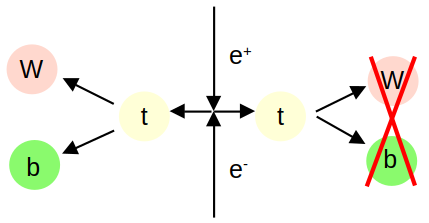}
  \caption{A schematic diagram of pair-production of top quarks at $e^+e^-$ colliders, with one top quark decaying into a SM-like channel and the other top
    quark decaying into a rare non SM-like channel.}
\label{fig:topveto}
\end{center}
\end{figure}

Here we propose to apply a similar strategy to the $t\bar t$ system in $e^+e^-\to t\,\bar{t}$, by tagging either the top or the anti-top quark as a SM top
(anti-top) decaying into $Wb$, and studying the recoil system
without making any assumption on its decay mode.
One could then build up a veto procedure to be applied on the second top system aiming to subtract all SM top pairs from the experimental sample. The final sample would then contain, besides the single tagged SM top
quark, any exotic top-decay event passing the SM top veto, as depicted in Fig.~\ref{fig:topveto}. The corresponding fraction of events with respect to the total sample would give an inclusive measurement of the {\it exotic top branching ratio} ${\rm BR}^t_{\rm exo}$. In the ideal case of perfect experimental resolution,  starting from $\mathcal{O} (10^6)$ top pairs (as foreseen at the FCC-$ee$), one could then probe down to ${\rm BR}^t_{\rm exo}\sim 10^{-5}$ in a model-independent way. 

The potential of this approach has to be compared with the well-known method to measure inclusively the exotic top width in $e^+e^-$  collisions, which consists of a precise
determination of the top width $\Gamma_t$ in $t\,\bar t$ production at threshold. Projections on the accuracy on $\Gamma_t$
reachable at the FCC-$ee$ correspond to probing the effect of a ${\rm BR}^{t}_{\rm exo}\;$ of the order of  $10^{-2}$~\cite{Defranchis:2025auz}. The present accuracy on  $\Gamma_t$ at the LHC~\cite{ParticleDataGroup:2024cfk}
would give an upper bound  around $0.13$ on the same quantity. On the other hand,  theoretical uncertainties due to missing higher orders on the SM top width would match a contribution from  ${\rm BR}^{t}_{\rm exo}$ of around $1.5 \times 10^{-2}$~\cite{Yan:2024hbz}\footnote{For a detailed discussion on a model-dependent approach to determine limits on the branching ratios of rare SM top decays, see~\cite{dEnterria:2023wjq}.}.

In case the event statistics and realistic experimental resolutions in 
top-recoil studies were not sufficient to detect a possible
${\rm BR}^{t}_{\rm exo}$ component, one could aim at just putting a
bound on ${\rm BR}^{t}_{\rm exo}$. To achieve this goal,  we start by  focusing on top {\it hadronic} decays which enhance  the control of kinematics in the $t\bar t$ reconstruction. Therefore, the potential of the proposed method will crucially depend on the final performance of the actual detector in tagging b-jets and in reconstructing the top hadronic systems.

We defer the analysis of detector effects in the reconstruction of the top system to subsequent studies, which involve $b$-tagging efficiencies as well,
and limit ourselves  to elaborate the efficiencies relying on 
Monte Carlo simulations and phenomenological reconstruction of SM $t\bar t$ hadronic events. 
We will therefore assume that the accuracy with which the above sample can be reconstructed will be directly related to the sensitivity with which  one can detect 
 inclusively possible BSM contributions in top decays, which will 
  survive a SM-top veto on the top-recoil system.
  
  Needless to say, the sensitivity we obtain corresponds to an ideal approximation that
will certainly be degraded by the inclusion of detection effects,
$b$-tagging,  and by the
 accuracy of the actual SM-top veto procedure.


This paper is organized as follows. In Section~\ref{sec:extract}
we present the results of the simulation of $t\bar t$ production
and decay in the SM all-jet channel in $e^+e^-$ collisions 
 at $\sqrt{s}=365$~GeV (as foreseen for  
one of the FCC-$ee$ energy stages above $t\bar t$ threshold),  showing
that the signal cross section is much
higher than the main SM backgrounds, which should allow a quite easy
extraction of $t\bar t$ events in future experiments.
In Section~\ref{sec:method}, we present a method to achieve the
maximum efficiency in reconstructing top quarks decaying in the SM fully-hadronic channel, presenting two different variants. Correspondingly,
we discuss the potential to bound 
the branching ratios for exotic top decays
inclusively. In
Section~\ref{sec:con} we provide the summary and make a few concluding remarks.
In Appendix ~\ref{sec:BSM}, we present an example of exclusive search
for a particular exotic top decay mode, and show how
this signal can be discriminated from the SM and 
manifest at the FCC-$ee$, following the strategies discussed in the paper.

\section{Simulation of \texorpdfstring{$t \bar{t}$}{ttbar} events and backgrounds at the FCC-\texorpdfstring{$ee$}{ee}}
\label{sec:extract}

As discussed in the introduction, 
this paper investigates the sensitivity of the \mbox{FCC-$ee$} 
to exotic top decays. To achieve this goal, it is essential to determine
the SM $t\bar t$ event reconstruction efficiency.
In order to deal with the highest possible cross section and
avoid final states with missing energy due to neutrinos, we 
study SM $t\bar t$ events in the all-hadron channel, namely the process:
\begin{equation}\label{allj}
  e^+e^-\to t \;(\to bW^+\to b jj) \;\;\bar t \;
  (\to \bar bW^-\!\to \bar b jj).
\end{equation}
The final state of the process in Eq.~(\ref{allj}) is then given by two
$b$-flavoured jets and four light jets ($jj$),
initiated by the $W$ decay products.
The same final state is also produced 
by several
other SM processes.
Below, we will consider the following backgrounds, which
have the highest cross sections:
\begin{enumerate}
    \item $e^+ e^- \to t \bar{b} W^-,\  t \to b W^+ , W^\pm \to j j$, where $\bar{b}$ and $W^-$ do not originate from $\bar{t}$;
    \item $e^+ e^- \to \bar{t} b W^+,\  \bar{t} \to \bar{b} W^- , W^\pm \to jj$, where $b$ and $W^+$ do not originate from $t$;
    \item $e^+ e^- \to b W^+ \bar{b} W^-,\  W^\pm \to j j$, where $b$, $\bar b$
      and $W^\pm$ do not originate from $t$ and $\bar t$.
\end{enumerate}
The $t\,\bar t$ events as well as the backgrounds are simulated at leading order
by means of the \textsc{MadGraph} code \cite{Alwall:2011uj}, matched to 
\textsc{PYTHIA} \cite{Sjostrand:2014zea} for parton showers and hadronization.
We set the top mass to $m_t$=173.2~GeV and
the $e^+e^-$ centre-of-mass energy to $\sqrt{s}=365$~GeV, about 20 GeV above
threshold. All other Monte Carlo parameters are set to the default values.
In this setup the cross section for $t\bar t$ production reads:
$\sigma(t\bar t)\simeq 0.18$~pb.
We cluster jets according to the $k_T$ (or Durham) algorithm for $e^+e^-$
annihilation in the $E$ recombination scheme
\cite{Catani:1991hj}, as embedded in the \textsc{Fastjet} code \cite{Cacciari:2011ma},
and require our events to have two $b$-jets and four light jets, for a total
of six jets.

In principle, accounting for next-to-leading order (NLO) corrections
in the narrow-width approximation (which neglects the interference between
top production and decay) is straightforward in the \textsc{MC@NLO}~\cite{Frixione:2002ik} or
\textsc{POWHEG}~\cite{Nason:2004rx, Frixione:2007vw, Alioli:2010xd} framework. However, including higher-order  effects is beyond the scopes of this
paper, since we are mostly interested in determining the order of magnitude
of the sensitivity of future $e^+e^-$ colliders to exotic top decays, rather then getting the most accurate computation in a given model.  We will defer to future work  the implementation of NLO corrections 
and, as discussed in the introduction, of $b$-tagging and detector effects, e.g., as in the \textsc{DELPHES}~\cite{deFavereau:2013fsa} framework.

\begin{figure}[htbp]
\begin{center}
  \includegraphics[height=0.3\textheight]{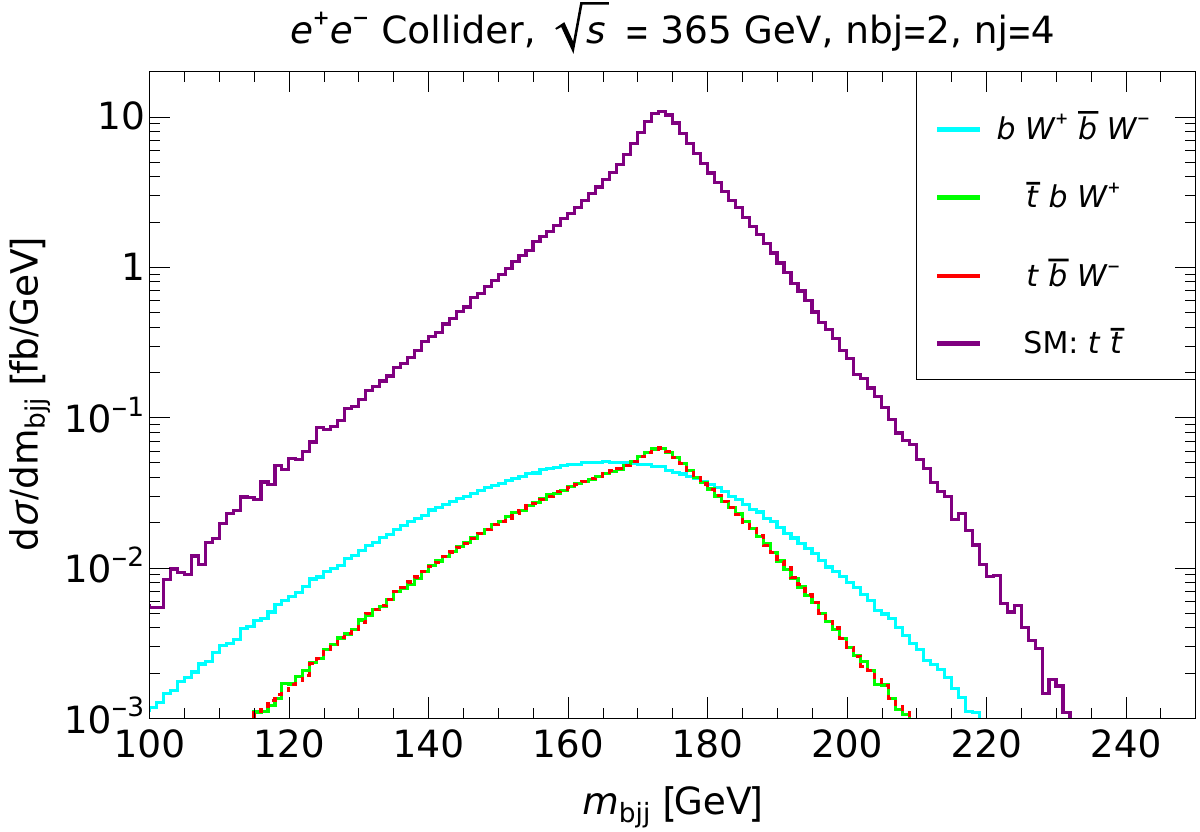}
\caption{Distribution of $m_{bjj}$ for $t\bar t$ events and the main SM background. These spectra show that the
  $t\bar t$ signal highly dominates over backgrounds across the entire 
  $m_{bjj}$ spectrum.}
\label{fig:thbw}
\end{center}
\end{figure}

In Fig.~\ref{fig:thbw} we present the distribution of the
invariant mass $m_{bjj}$ of one $b$-jet and 2 light jets for $t\bar t$ events (purple histograms), and the above-mentioned three SM backgrounds (red, green and cyan, respectively), on logarithmic scales.
When pairing $b$- and light-flavoured jets, we
choose the $bjj$ combination which is the closest to the top mass. 
More details on the clustering and the strategy to combine $b$-flavoured and light jets will be given in the next section.
Fig.~\ref{fig:thbw} shows that 
the cross section of the $t\bar t$
signal 
is much higher than that of
the background processes across the entire $m_{bjj}$  interval.
In particular, around the peak $m_{bjj}\simeq m_t$, the
$t\bar t$ spectrum is more than two order of magnitudes larger than the
backgrounds. As for the backgrounds, the $bWbW$ one yields a higher rate
than the $tbW$ ones in the full range, with the exception of the region around the peak. As expected, the $\bar tbW^+$ and $t\bar bW^-$ spectra
are statistically equivalent.
The conclusion of the comparison in Fig.~\ref{fig:thbw} is that it
is quite straightforward to discriminate the $t\bar t$ signal from the
backgrounds at $e^+e^-$ colliders with $\sqrt{s}\sim 365$~GeV in the all-jet channel.


Now that we have shown that the $t\bar t$ signal in the all-jet channel
has roughly no background contamination, we focus on reconstructing the top mass
 from the top hadronic decay products with the highest possible efficiency.
In the following section, we consider two slightly different algorithms to achieve this goal.

\section{Reconstructing top quarks}
\label{sec:method}

In this section we determine the efficiency to reconstruct
top quarks at the FCC-$ee$ through their decay products. In fact, as explained
above, in order to set any bound on the sensitivity to exotic decays, it is
of paramount importance to have under control the capability of reconstructing
SM final states. As in the previous section, our investigation will be
carried out at leading order (LO), with the \textsc{MadGraph} matrix-element
generator matched to \textsc{PYTHIA}, and we will  deal with the all-hadronic
top decay channel, as in Eq.~(\ref{allj}).
Jets originated from top and $W$ decays will be clustered into
six jets according to the Durham algorithm in the $E$ recombination scheme~\cite{Catani:1991hj}.

Once samples of events with two $b$- and four light jets are simulated,
the next step consists in selecting the correct combination of jets
to identify the two decaying top quarks. We will scrutinize two different
methods to achieve our goal, and eventually choose the one which
leads to the highest reconstruction efficiency.

\subsection{\texorpdfstring{$\chi^2$}{chisq} definition and minimization}
\label{sec:chi}

Since we are dealing with the all-jet channel, ideally one expects that
the right combination of light jets should yield the 
$W$ bosons, while such light jets plus the $b$-flavoured jet coming
from the same decaying top should reconstruct the top quark.
In other words, the invariant mass of the two ``right'' light jets should
give $m_W$ and the one of the correct combination of two light jets plus
a $b$-jet should yield $m_t$.

As pointed out in the introduction, in our investigation a role will be played
by the so-called {\it recoil mass}, 
$ m_{\rm rec}$.
In $t\bar t$ events, assuming that 
only one of the
two top quarks is tagged, the mass of the recoiling system is defined as:
\begin{equation}
    m_{\rm rec}^2 = (Q-p_t)^2=s + m_t^2 - 2\ \sqrt{s}\ E_t\;,
    \label{eq:mrec}
\end{equation}
and should match 
$m_t$.
In Eq.~(\ref{eq:mrec}), $\;Q$ is the  $e^+ \!+ e^-$ four-momentum, i.e.
$Q=(\sqrt{s},0,0,0)$ in the laboratory frame, $p_t$ is the four momentum of the tagged top quark
and $E_t$ its energy in the laboratory frame.
One can note that, although $m_{\rm rec}$ refers to the recoiling system,
it is expressed only in terms of the reconstructed particle and the
$e^+ \!+ e^-$ four momentum, while it contains no information on either
nature or kinematics (apart from its mass $m_{\rm rec}$) of the particle which recoils against the tagged one.
This makes the recoil mass particularly suitable to 
explore in a model independent way any possible exotic process
occurring in the system recoiling against the top~\footnote{A similar use of the
recoil mass is made in the $e^+e^-\to ZH$ associated production, 
where one can constrain the Higgs boson
by studying the system recoiling against the $Z$\cite{Ito:2009jf,Lohmann:2007ty}.}.

\begin{figure}[ht!] 
\hskip 0.8cm
  \begin{subfigure}[h]{0.45\textwidth}
  \centering
  \hskip -0.8cm
  \includegraphics[width=1.\linewidth]{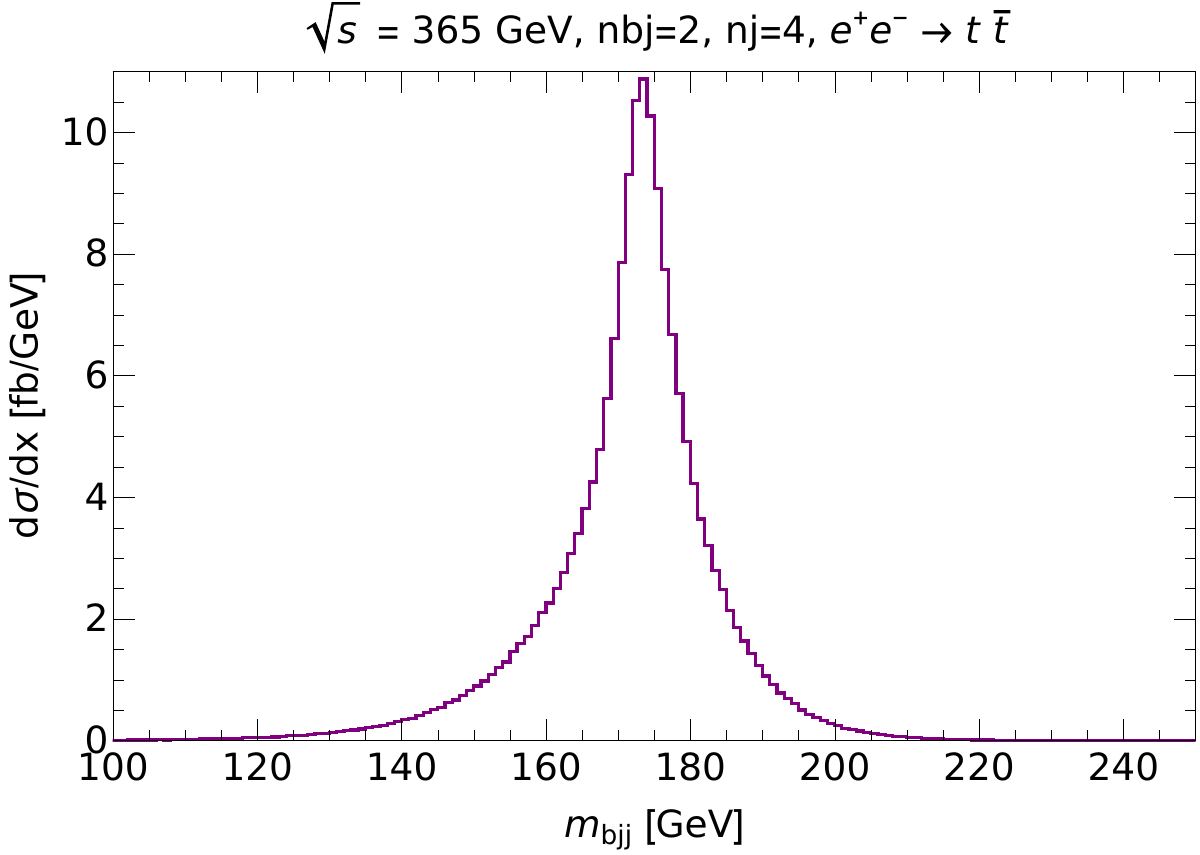}
  \caption{}
  \label{fig:topa}
\end{subfigure}%
  \begin{subfigure}[h]{0.45\textwidth}
    \centering
    \includegraphics[width=1.\linewidth]{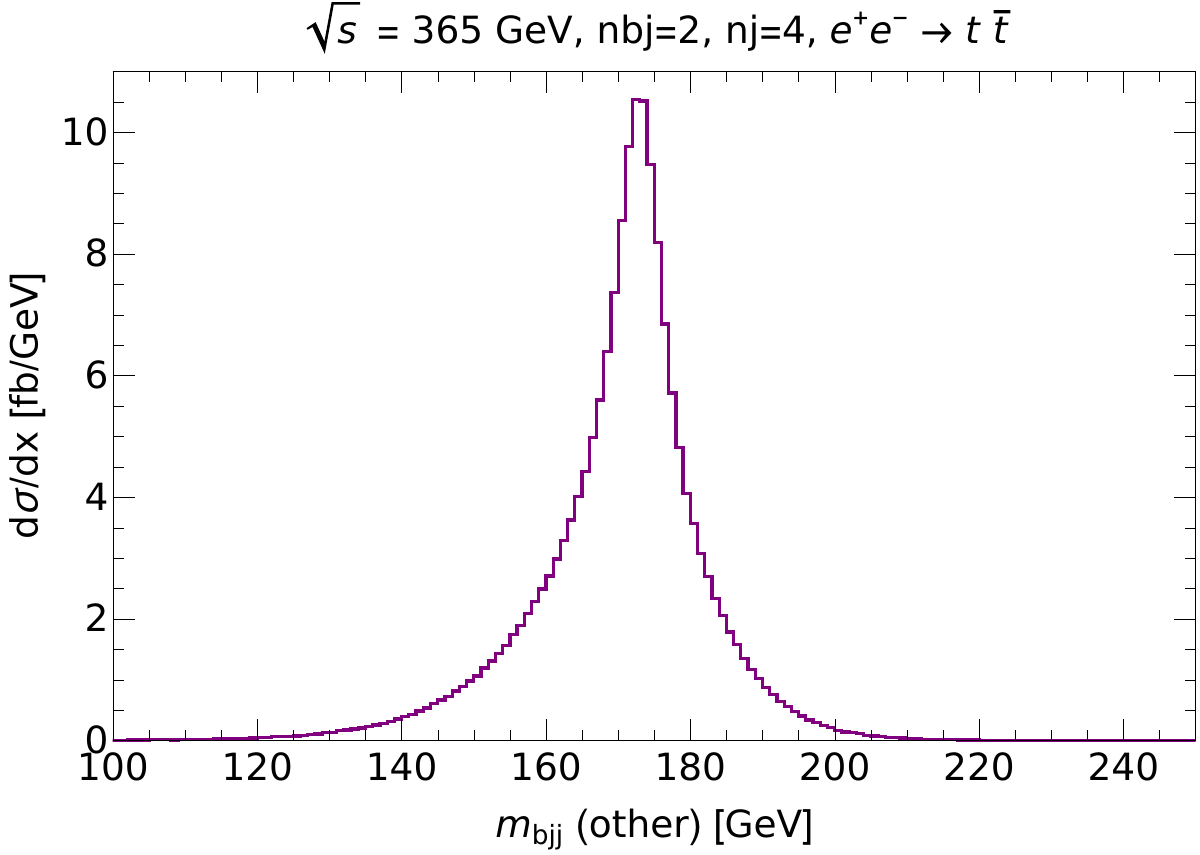} 
    \caption{}
    \label{fig:invmbjjothera}
  \end{subfigure}
  \begin{subfigure}[h]{0.45\textwidth}
   \hspace*{0.3 cm}
    \centering
    \includegraphics[width=1.\linewidth]{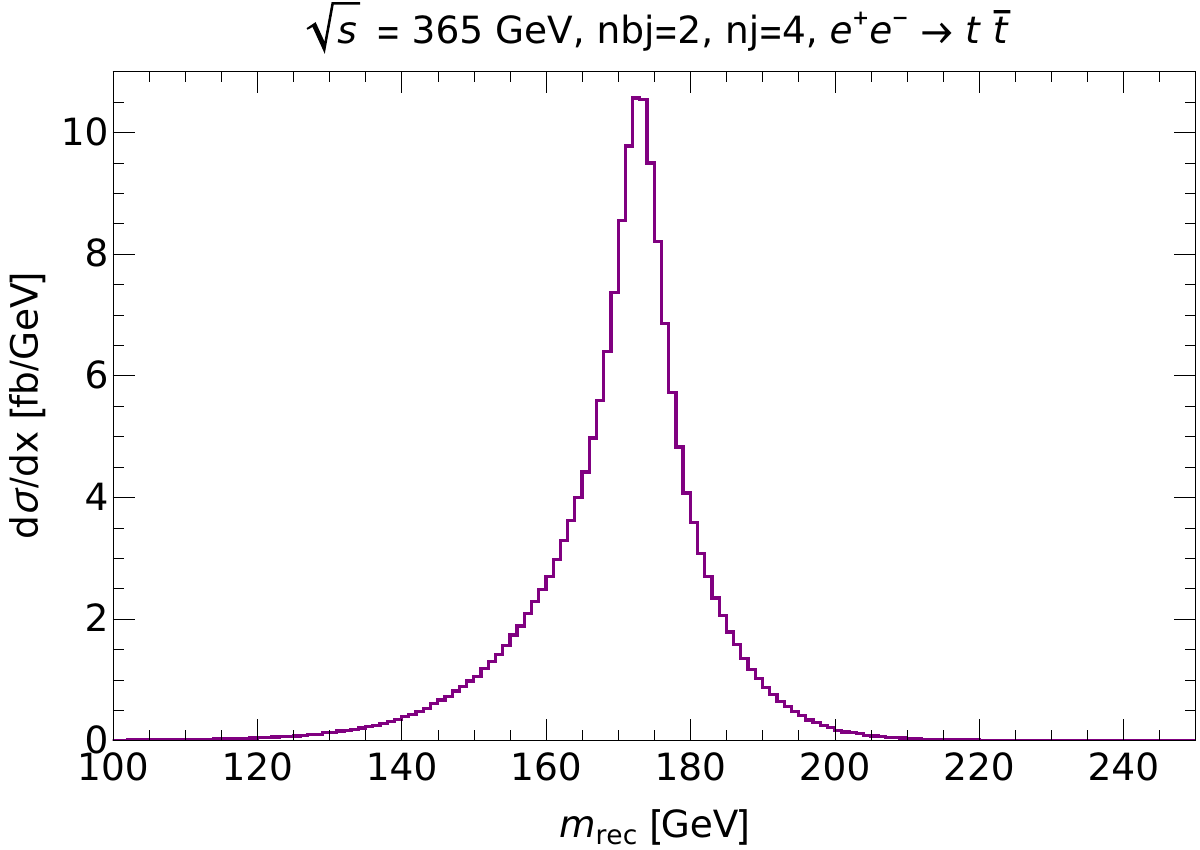} 
    \caption{} 
    \label{fig:mreca}
  \end{subfigure}
  \begin{subfigure}[h]{0.47\textwidth}
   \hspace*{0.5 cm}
    \centering
    \includegraphics[width=1.\linewidth]{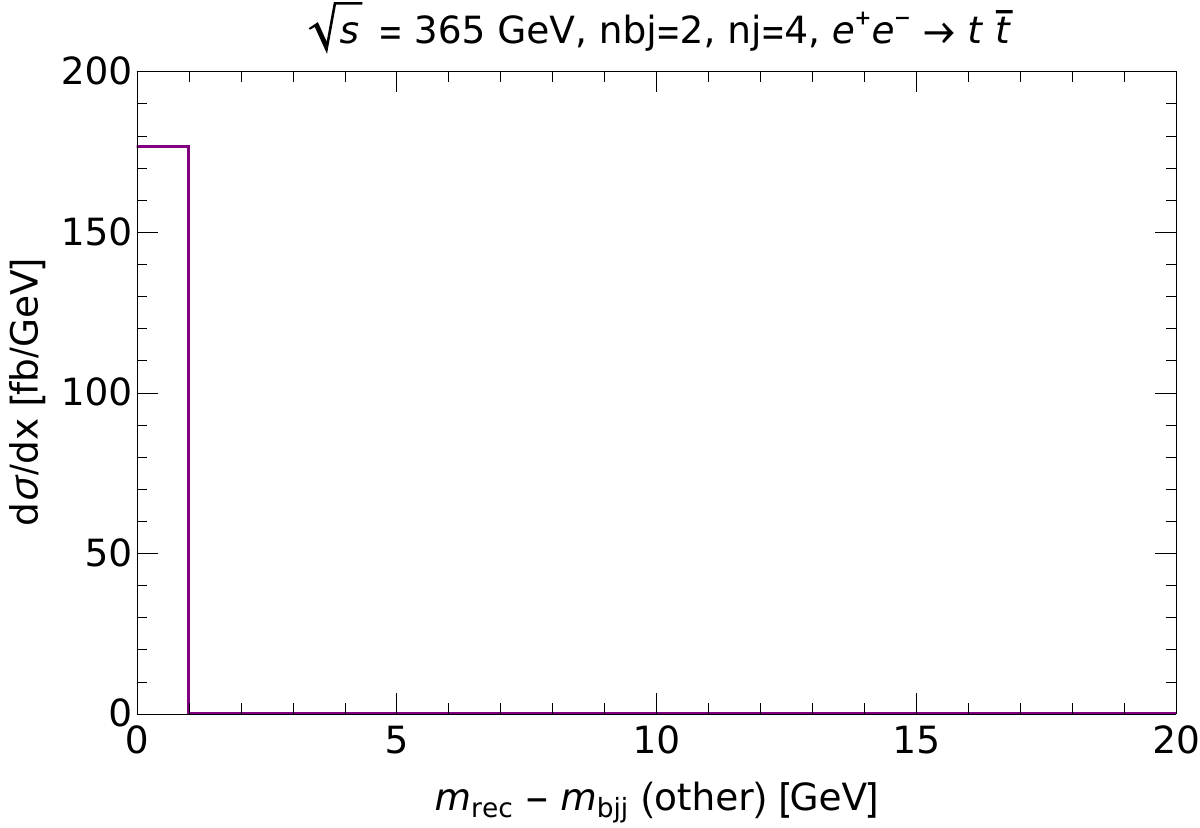} 
    \caption{} 
    \label{fig:diffreca}
  \end{subfigure} 
\vspace*{-0.1in}
\caption{Distributions of (a) invariant mass of one $b$-jet and two
  light jets for which the $\chi^2$ is minimum, (b) invariant mass of the leftover $b$-jet and two light jets (``$m_{bjj}$ other''), (c) recoil mass calculated
  with respect to the $bjj$ combination minimizing the $\chi^2$,  (d)
  difference between the recoil mass and ``$m_{bjj}$ other''.}
 \label{fig:minchi}
\end{figure}

With these requirements to be met, we found that an optimal way to select the correct combinations of light and $b$-jets to reconstruct the top
quark
consists of minimizing the
 following $\chi^2$:
 \begin{equation}\label{chi2}
   \chi^2= (m_{jj}-m_W)^2 + (m_{bjj}-m_t)^2 + (m_{\rm rec}-m_t)^2.
   \end{equation}
 In other words, we select the pair of light jets whose invariant
 mass is near the $W$ mass, and the combination of $b$- and light jets, as well as
 the recoil system, whose invariant mass is as close as possible  to $m_t$.
 The results after the $\chi^2$ minimization are presented    
in Fig.~\ref{fig:minchi}.
In particular, in Fig.~\ref{fig:topa} we plot the invariant mass $m_{bjj}$
of the $bjj$ combination
which minimizes the $\chi^2$ in Eq.~(\ref{chi2}) (presented also in Fig.~\ref{fig:thbw}, {\it purple} histogram, but on a logarithmic  scale). The
$m_{bjj}$ spectrum has the typical
rise-and-fall behaviour, is substantial for $120~{\rm GeV}<m_{bjj}<220~{\rm GeV}$
and is peaked around $m_t=173.2$~GeV. Assuming that one has tagged
one top minimizing the $\chi^2$, Fig.~\ref{fig:invmbjjothera} presents the 
invariant mass of the other two light jets along with the other $b$-jet,
labelled as ``$m_{bjj}$ other''.   
Fig.~\ref{fig:mreca} shows the recoil mass, calculated according to
Eq.~(\ref{eq:mrec}). The three spectra shown in
Figs.~\ref{fig:topa}--\ref{fig:mreca} look roughly the same, which shows  the reliability of our method based on the minimization
of  $\chi^2$ in Eq.~(\ref{chi2}). In fact, after combining the jets in the
``right'' manner even the mass of the recoil system  matches the top mass.
As a further check,
Fig.~\ref{fig:diffreca} shows the difference between the
recoil mass and ``$m_{bjj}$ other''.  The resulting spectrum is sharply
peaked around zero, which still confirms the goodness of the
use of  $\chi^2$ in Eq.~(\ref{chi2}).

In order to use the results in Figs.~\ref{fig:topa}--\ref{fig:diffreca} 
to eventually bound exotic top decays, we need to assess the
efficiency to reconstruct SM top events in our set up.
We first observe that, when running the Durham algorithm, a small fraction
of events (about 2\%) is not capable of yielding the six required
jets. In fact, one may have events where the two $B$-hadrons in top decays
are so close in phase space that get clustered in one single $b$-jet. 
Likewise, two light jets or one light and one $b$-jet may mix up.
Furthermore, though being quite rare, one may always have
$g\to b\bar b$ splittings in the shower, which increases the number of
$b$-jets.

Having observed this, we focus our investigation on the events which do have
six jets and, as can be understood from the histograms in
Fig.~\ref{fig:minchi}, most SM events occur in a range 
of a few dozens of GeV around $m_t$.
Setting as a working assumption  $\Delta m_{bjj}\simeq 50$~GeV,
we found that about 99.5\% of the event with six jets
has a $m_{bjj}$ invariant mass
in the range:
\begin{equation}\label{deltam}
  m_t-50~{\rm GeV}<m_{bjj}<m_t+50~{\rm GeV}.
  \end{equation}
The small discrepancy, amounting
to about 0.5\%, which is the fraction of {\it non-reconstructed events}, 
can be interpreted as an inclusive upper bound on
the branching ratio of exotic top decays
within our reconstruction method. We underline that this
limit has been set in a completely
inclusive and model-independent way, as it is based on the sole reconstruction
of one SM top quark and on the efficiency to identify the $m_{bjj}$
invariant mass as a top mass in a given range. Furthermore,
we checked that by varying $\Delta m_{bjj}$ by few dozens of
GeV, e.g. $\Delta m_{bjj}\sim\, $40 or 60 GeV,
the fraction of non-reconstructed events is quite stable.

As a whole,
considering that this is a first attempt to address the $t\bar t$ events
reconstruction as a chance to bound inclusively BSM decay
modes, this result sounds quite promising. Nevertheless we have tried
to further improve the 99.5\% efficiency, as will be detailed in the
following subsection.

\subsection{Alternative \texorpdfstring{$\chi'\,^2$}{chipsq} definition and minimization}

\begin{figure}[h!]
\begin{center}
  \includegraphics[height=0.3\textheight]{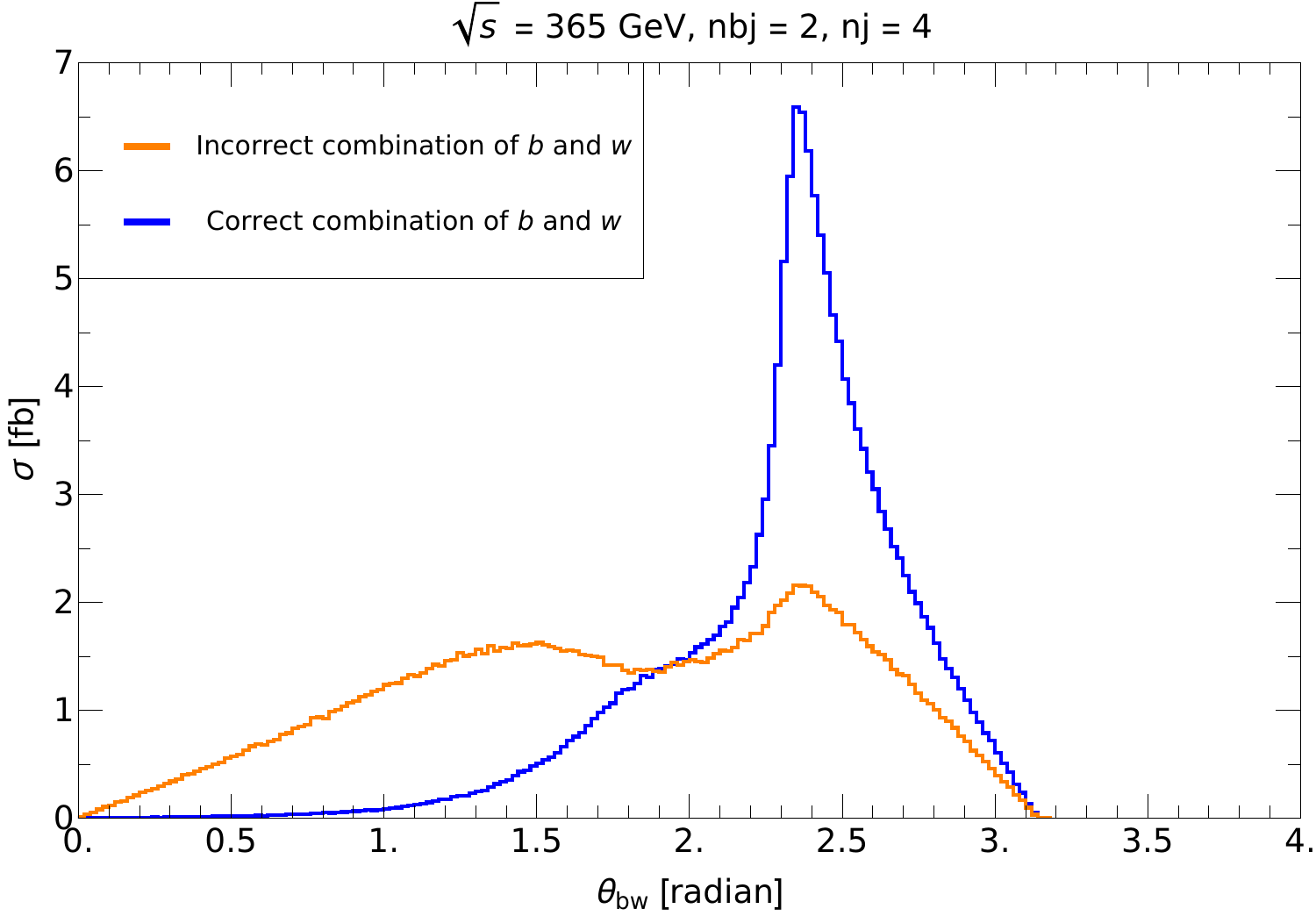}
  \caption{Distribution of angular distance between $b$-jet and $W^{\pm}$ in
    top decays, where $b$ and $W$ are paired correctly (blue) or not (orange).}
\label{fig:thbw2}
\end{center}
\end{figure}

In order to investigate whether one can further improve the efficiency 
of top-quark identification and then be sensitive to even lower
${\rm BR}^t_{\rm exo}$ values, one may try to add to the $\chi^2$ definition
in Eq.~(\ref{chi2}) extra terms. In particular,
we explored the angle between 
$b$-jet and $W$, assuming that the latter can be reconstructed
by using the two light jets $jj$. Fig.~\ref{fig:thbw2} presents
the spectrum of the angle $\theta_{bW}$, according to whether $b$-jet and
$W$ are paired correctly (blue histogram) or not (orange).
The two spectra are remarkably different; in particular the one corresponding
to the correct combination is peaked at $\theta_{bW}\sim$2.35 radians at $\sqrt s = 365$~GeV. One may therefore think of a modified $\chi'^2$ definition,  including this
feature of the $\theta_{bW}$ angular distribution, as follows:
\begin{equation}\label{chiprime}
  \chi'^2=(m_{jj}-m_W)^2 + (m_{bjj}-m_t)^2 + (m_{\rm rec}-m_t)^2 + (\theta_{bW}-2.35)^2.\end{equation}
  We repeated the same analysis made for the previous $\chi^2$ definition and
  found comparable results, namely spectra consistent with those
  in Figs.~\ref{fig:topa}--\ref{fig:diffreca}, and 99.5\% of events
  with \mbox{$m_t-50~{\rm GeV} <m_{bjj}<m_t+50~{\rm GeV}$}. 
  We conclude that including the non-trivial $\theta_{bW}$ angular features
  in the analysis does not improve sensibly the results.

\section{Summary and Outlook}
\label{sec:con}

Future $e^+e^-$ colliders, such as the FCC-$ee$ or linear colliders,
will allow measurements
of the top-quark properties with  unprecedented precision, thanks to the
clean environment of electron-positron annihilation.
In particular, it will be possible to set inclusive bounds on exotic top decay branching ratios  
independently of the specific new physics model, which, on the contrary,
is not realistic at   the LHC.

We  simulated $t\bar t$ production at the FCC-$ee$  above threshold, 
at $\sqrt{s}=365$~GeV,
concentrating
on the all-hadron channel, and found that it is relatively straightforward
discriminating events with top pairs from the backgrounds.
Then we set up a simple strategy, based on the $k_T$ jet-clustering algorithm
and the use of the recoil mass, to reconstruct top quarks in $e^+e^-$ collisions,
and obtained an efficiency of about 99.5\%. 
The remaining 0.5\% expresses 
the sensitivity of the FCC-$ee$  to possible exotic branching ratios,
which we were able to set inclusively and
without making any assumption on possible BSM scenarios.

We stress 
that our phenomenological 
estimate of the reconstruction efficiencies corresponds to an {\it ideal} picture that will be degraded by the inclusion of detector effects such as $b$-tagging and other detector efficiencies.
With this caveat, our result shows a higher potential to constrain
inclusively possible excesses in exotic top BR's 
than the bound of about 1\% obtained instead from 
 the expected accuracy on the $\Gamma_t$ measurement at the FCC-$ee$ at threshold.
The inclusion of NLO corrections, e.g. in the narrow-width approximation,
as well as detector effects, which are compelling to give a more
realistic and precise
estimate of the sensitivity to BSM decays, is deferred to future work.

In summary, we believe that this investigation may well be seen as a useful
starting point to address top phenomenology above threshold
in the clean environment of $e^+e^-$ colliders from an inclusive and
model-independent perspective. We hope to return in the future with
further explorations which may help  guiding and motivating experimental analyses.

\section*{Acknowledgments} 
We acknowledge discussions on the FCC-$ee$ with Michele Selvaggi and Xunwu Zuo.

\appendix
\section{Impact of a non-SM top decay on the simulation}
\label{sec:BSM}

In this appendix we apply our method, which has been developed in a
completely model-independent manner, to a specific BSM top decay.
In the introduction, we mentioned a few BSM top decays  which have so far been investigated mostly at the LHC.
Hereafter, for illustrative purposes, we shall investigate the 
channel with a charged Higgs boson and a bottom quark, followed by the Higgs decay 
into a neutrino and a tau lepton, eventually decaying hadronically:
\begin{equation}\label{hb}
  t\to H^+b\to(\tau^+\nu_\tau)b\to (jj\bar\nu_\tau\nu_\tau)b.
  \end{equation}

\begin{figure}[ht!]
\centering
\begin{subfigure}[h]{0.45\textwidth}
  \centering
  \includegraphics[width=1\linewidth]{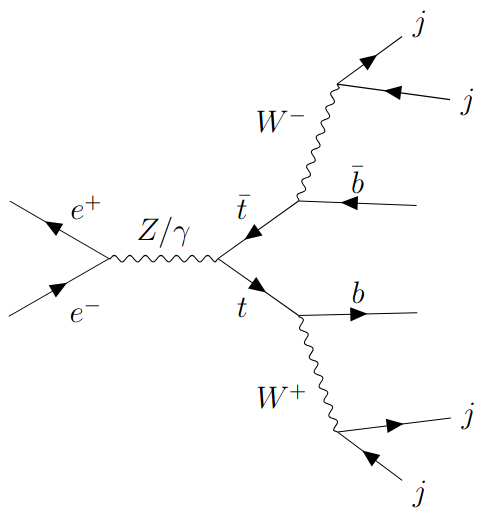}
  \caption{}
\label{fig:ttbar}
\end{subfigure}%
\begin{subfigure}[h]{0.45\textwidth}
  \centering
  \includegraphics[width=1\linewidth]{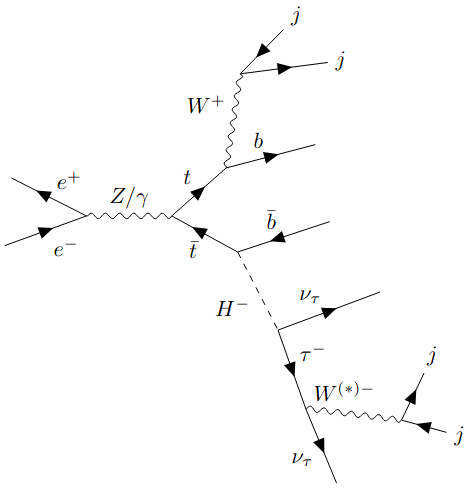}
  \caption{}
\label{fig:ttbarh}
\end{subfigure}
\vspace*{-0.1in}
\caption{Feynman diagrams for $t\bar t$ production
  and decay in $e^+e^-$ annihilation in (a)  SM all-jet channel
  and (b) one $W$ replaced by a charged-Higgs, followed by
  $H^\pm$ decays as discussed in the text.}
 \label{fig:feyn}
\end{figure}

The top decay into $H^+b$
is kinematically similar to the SM $W^+b$ channel. The $\tau$ decay occurs  via a virtual
$W$ whose decay into light jets mimics the one of the real $W$ in top decay, but with lighter $m_{jj}$.
Another difference of the final states
yielded by the decay chain
(\ref{hb}) with respect to the all-jet SM one is the presence of missing
energy due to the tauonic neutrinos.
The SM and BSM processes which we wish to compare are
presented in Fig.~\ref{fig:ttbar} and Fig.~\ref{fig:ttbarh} respectively.

As above, we simulate such processes at
$\sqrt{s}=365$~GeV with \textsc{MadGraph} and \textsc{PYTHIA}, 
and cluster jets according to the Durham algorithm.
Though any consideration on specific new physics scenarios is beyond the
goal of this work, for practical purposes
$t\to bH^+$ decays are simulated according to the
type II seesaw model, implemented in  \textsc{MadGraph}
as in~\cite{Fuks_2020}, with a $H^+$ mass set to 
101 GeV.

\begin{figure}[ht!]
\hskip 0.8cm
  \begin{subfigure}[h]{0.45\textwidth}
  \centering
  \hskip -0.8cm
  \includegraphics[width=1.0\linewidth]{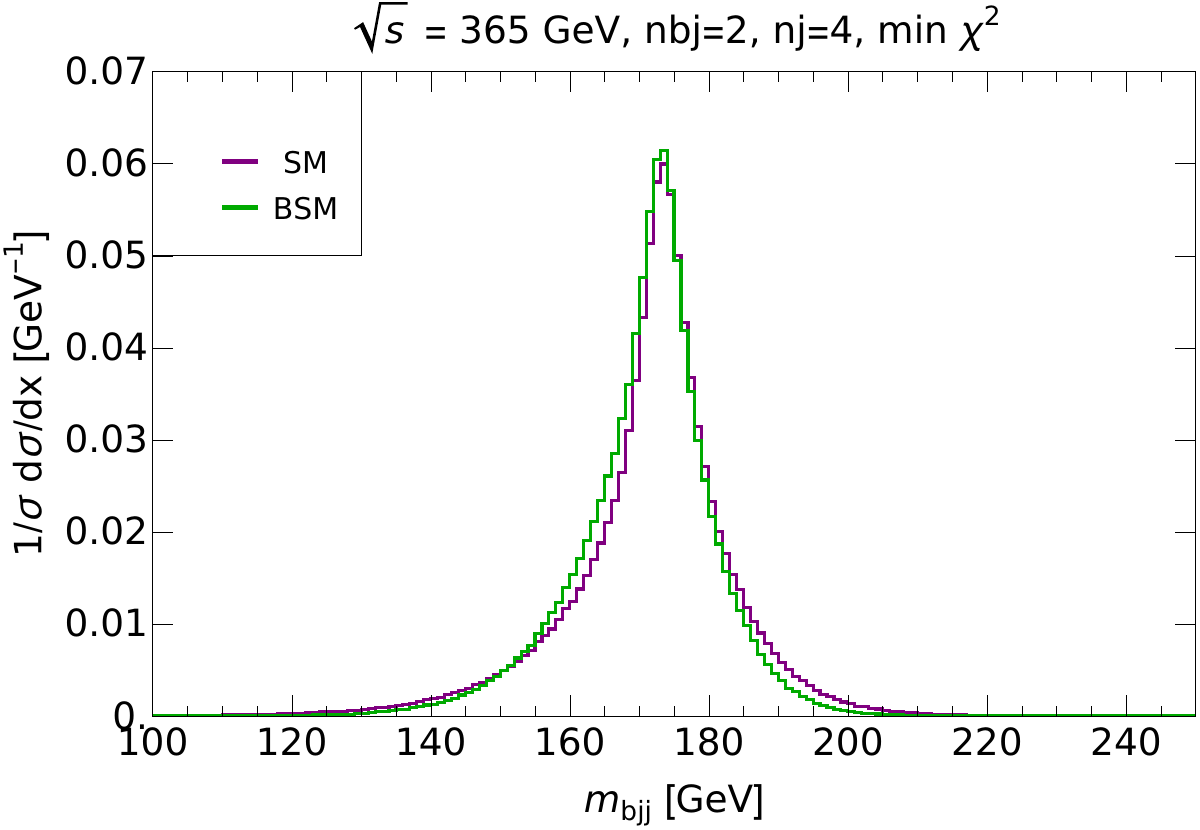}
  \caption{}
  \label{fig:topc}
\end{subfigure}%
  \begin{subfigure}[h]{0.45\textwidth}
    \centering
    \includegraphics[width=1.0\linewidth]{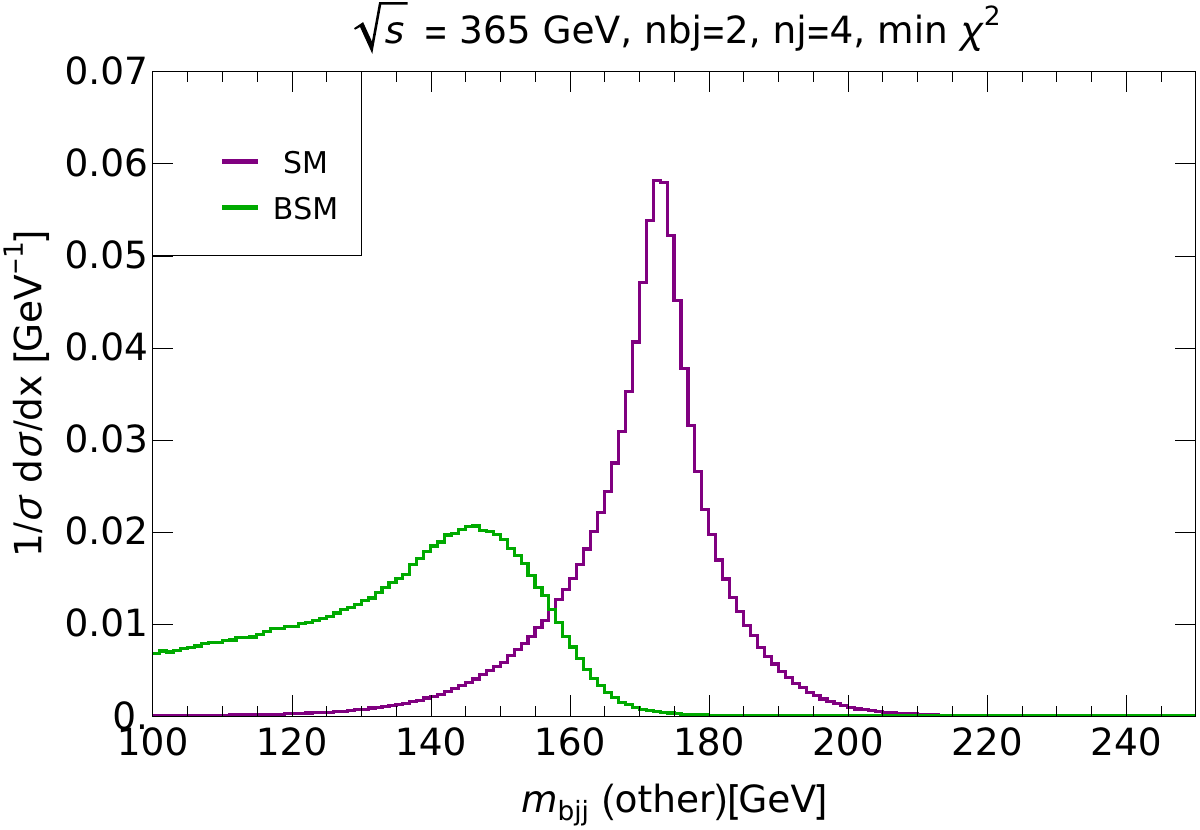} 
    \caption{}
    \label{fig:invmbjjotherc}
  \end{subfigure} 
  \begin{subfigure}[h]{0.45\textwidth}
  \hspace*{0.3 cm}
    \centering
    \includegraphics[width=1.\linewidth]{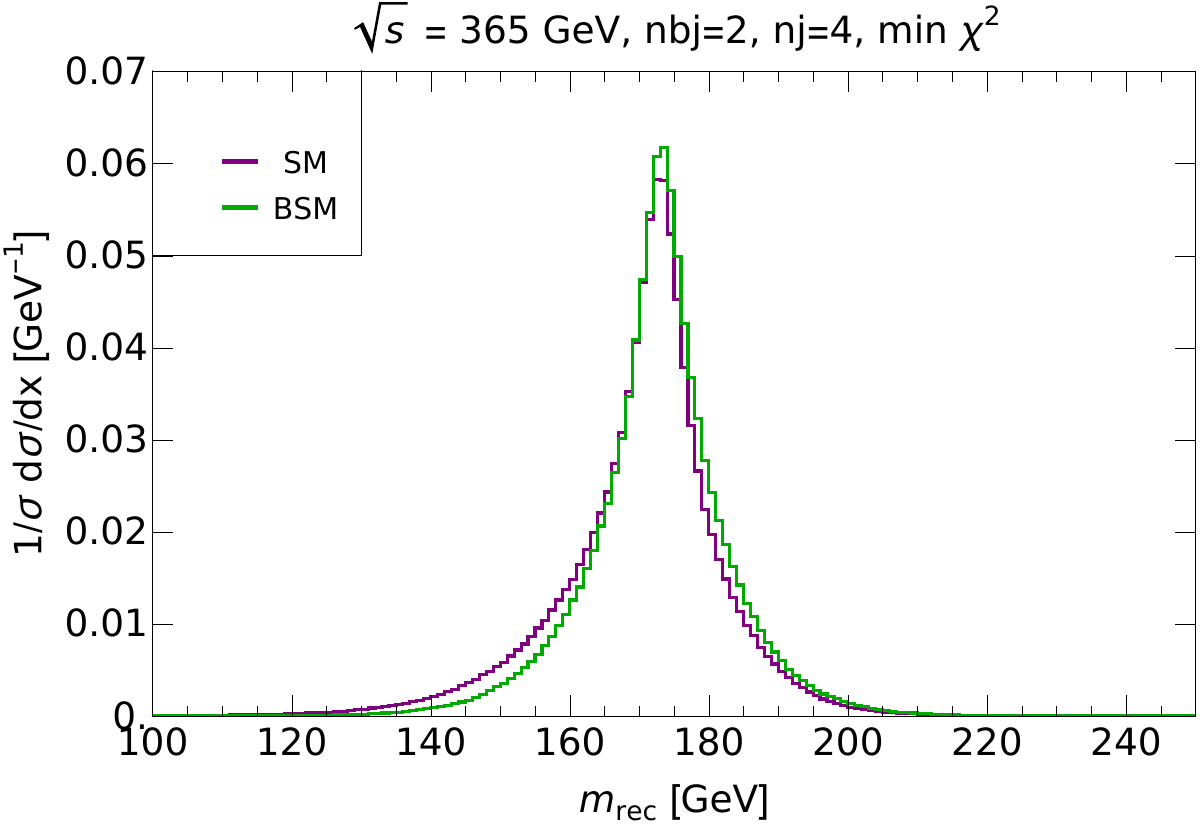} 
    \caption{} 
    \label{fig:mrecc}
  \end{subfigure}
  \begin{subfigure}[h]{0.46\textwidth}
  \hspace*{0.6 cm}
    \centering
    \includegraphics[width=1.\linewidth]{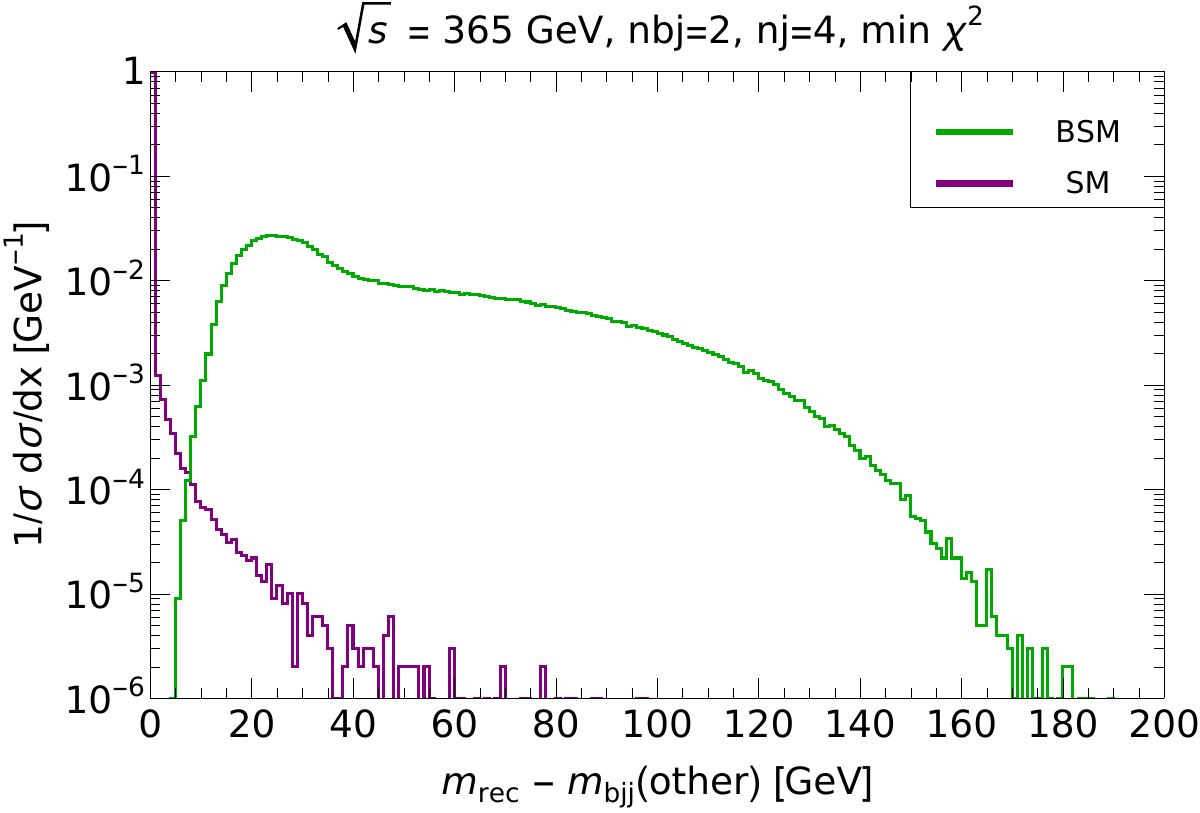} 
    \caption{} 
    \label{fig:diffrecc}
  \end{subfigure} 
\vspace*{-0.1in}
\caption{Normalized distribution of (a) invariant mass of one $b$-jet and two light jets for which the $\chi^2$ is minimum,
  (b) invariant mass of the remaining $b$- and light jets, (c) recoil mass with
  respect to the SM decaying top quark
  (d) difference between the recoil mass and ``$m_{bjj}$ other''.
 The distributions for the SM process (Fig.~\ref{fig:ttbar}) are  denoted by the purple curves and those for the BSM process (Fig.~\ref{fig:ttbarh}) are denoted by the green curves.}
 \label{fig:minchic}
\end{figure}

Bottom-flavoured and light jets are combined by minimizing the
$\chi^2$ defined as in Eq.~(\ref{chi2}). As before, we study $i)$~the invariant
mass $m_{bjj}$, $bjj$ being the jet combination which minimizes
the $\chi^2$ and should then yield the top mass, $ii)$ the so-called
``$m_{bjj}$ other'', which corresponds to the other $bjj$ combination, $iii)$ the
recoil mass, and $iv)$ the difference between recoil mass and ``$m_{bjj}$ other''.
Such plots are present in Fig.~\ref{fig:minchic} where the purple curve denotes the SM process while the green curve denotes the BSM process. Since the dynamics and
the kinematics of the $\bar t$ decay are different from the standard case,
some discrepancies are to be expected.
The normalized $m_{bjj}$ spectra in Fig.~\ref{fig:topc} have a similar
behaviour, but a visible shift of the BSM one towards larger $m_{bjj}$ values can well be seen, due to the different final state which yields a slightly different $m_{bjj}$ spectrum.
Furthermore, as expected, the distributions of ``$m_{bjj}$ other'',
which refer to the untagged top, are quite different, with the BSM one
exhibiting a broad peak around 150 GeV, vanishing above 180 GeV and
yielding many more events than the SM one at low invariant masses.
In fact, in the BSM process, ``$m_{bjj}$ other'' is not capable
of yielding the correct top mass
(due to missing energy from neutrinos), and  does not peak at $m_t$.
The recoil mass in Fig.~\ref{fig:mrecc}
  looks qualitatively similar to the $m_{bjj}$ ones, as it should 
  since it depends only on the kinematics of the tagged (SM) top
  quark.
Also, as expected, the slight shift of the BSM distribution is in the
opposite direction with respect to the one observed in $m_{bjj}$ in
Fig.~\ref{fig:topc}, i.e., towards lower values of the recoil mass. Nevertheless, both SM and BSM process peak at $m_{t} \sim 173.2$~GeV in Fig.~\ref{fig:topc} and \ref{fig:mrecc}, with almost all events exhibiting both 
$m_{bjj}$ and $m_{\rm rec}$ in the range between 140 and 220 GeV.
As observed in Section 3, 
one can expect a slight difference in the estimate of the efficiency of reconstructing the top mass if this mass window is changed.
This difference is anyhow negligibly small for a reasonable width
of such a window.
Finally, in Fig.~\ref{fig:diffrecc} we present 
(on a logarithmic scale) the difference between recoil-mass and
``$m_{bjj}$ other'' spectra. While the SM case
is again sharply peaked around zero, the BSM one has a very broad spectrum,
peaked at about 30 GeV, and 
substantial especially for small mass values. Once again, this feature of the
BSM curve in Fig.~\ref{fig:diffrecc} reflects the fact that we are subtracting
from the recoil mass (computed from the kinematics of a SM tagged top)
the  invariant mass corresponding to $\bar t\to H^-\bar b$, with $H^-\to
\tau^-\bar\nu_\tau$ characterized by the presence of missing momentum.


\begin{figure}[ht!]
\centering
  \includegraphics[width=0.7\linewidth]{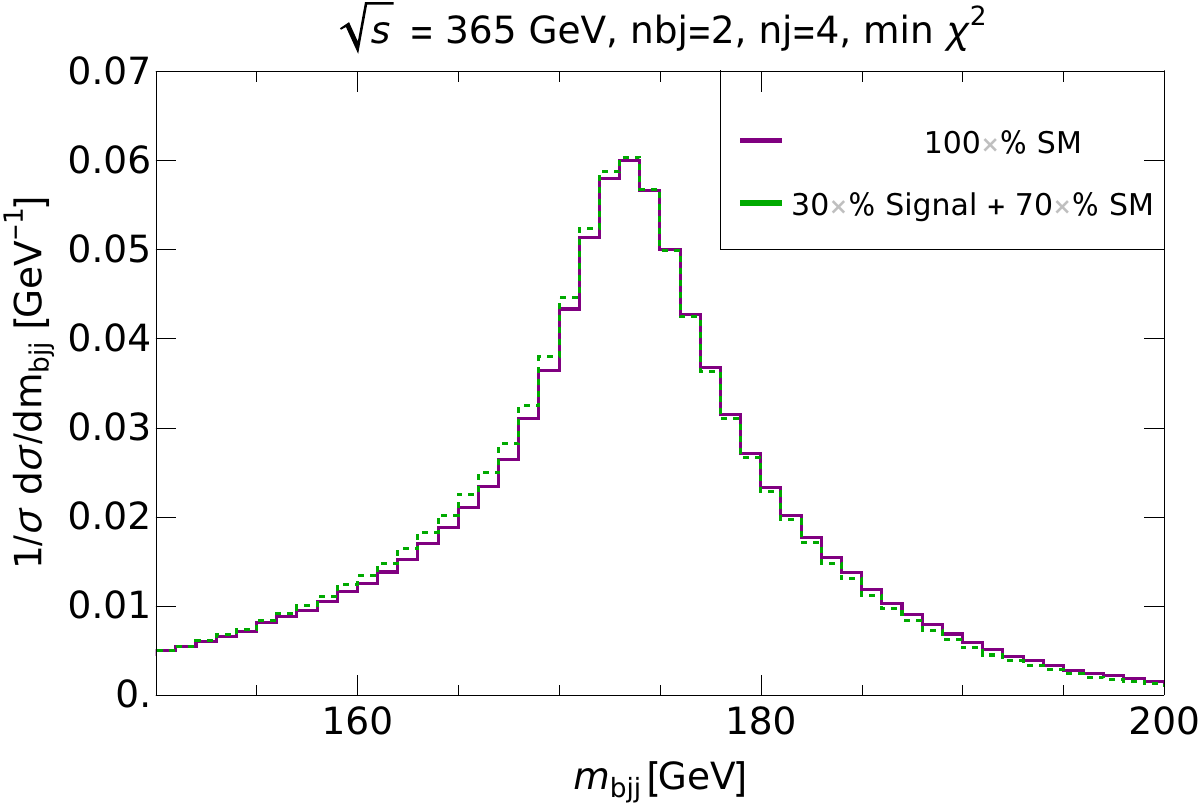}
\vspace*{-0.1in}
\caption{Normalized distribution of $m_{bjj}$
  for the fully SM process (purple curve) and a signal (green) where the
  contamination of the BSM $\;t\to bH^+$ decay is  30\%.}
\label{fig:wgtnormchi}
\end{figure}

For illustrative purposes, we wish now
to investigate how the $m_{bjj}$ distribution gets modified when the SM all-jet
channel is contaminated by a contribution given by the BSM process
in Eq.(\ref{hb}).
A quantitative estimate
of the $t\to H^+b$ branching ratio is model-dependent and  beyond
the scopes of this paper. 
Nevertheless, in order to see the  impact of this channel on the $m_{bjj}$ distribution, we just make the assumption that  the channel 
$t\to H^+b\ (H^+\to \tau^+\nu_\tau,\tau^+\to jj\bar\nu_\tau)$ contributes
to a  30\% fraction of the events.
Our results are presented in Fig.~\ref{fig:wgtnormchi}, where we compare
the sole SM normalized cross section with the one given by 70\% SM and 30\% BSM
contributions. As already
observed in Fig.~\ref{fig:minchic}, $t\to bH^+$ decays slightly shift the $m_{bjj}$ spectrum towards lower values.

Before concluding the appendix, we wish to stress that  in Figs.~\ref{fig:minchic} and \ref{fig:wgtnormchi} 
we
have plotted normalized
spectra, and the assumption of a 30\% BSM branching fraction
in Fig.~\ref{fig:wgtnormchi} has somewhat enhanced the BSM contribution
and the shift in the $m_{bjj}$ distribution. Any realistic assumption
on a possible BSM contamination of the SM cross section would
have a much milder effect, and would have no impact on  the conclusion of our analysis, which led to 
 a 0.5\% sensitivity of the FCC-$ee$ to bound exotic top decays
in a model-independent way.


\bibliographystyle{utphys}
\bibliography{reference}

\end{document}